\documentclass[runningheads]{svmult}

\usepackage{makeidx}   
\usepackage{graphicx}  
\usepackage{subeqnar}  
\usepackage{multicol}  
\usepackage{physprbb}  
\makeindex             

\begin{document}
\title*{Search and Congestion in Complex Networks}
\toctitle{Search and Congestion in Complex Networks}
%
%
\titlerunning{Search and Congestion in Complex Networks}
%
\author{Alex Arenas\inst{1}
\and Antonio Cabrales\inst{2}
\and Albert D{\'\i}az-Guilera\inst{3}
\and Roger Guimer{\`a}\inst{4}
\and Fernando Vega-Redondo\inst{5}
}
\authorrunning{Arenas, Cabrales, D{\'\i}az-Guilera, Guimer{\`a} and Vega-Redondo}
%
%
\institute{Departament d'Enginyeria Inform{\`a}tica, Universitat Rovira i Virgili, Tarragona
\and Departament d'Economia i Empresa, Universitat Pompeu Fabra, Barcelona
\and Departament de F{\'\i}sica Fonamental, Universitat de Barcelona, Barcelona
\and Departament d'Enginyeria Qu{\'\i}mica, Universitat Rovira i Virgili, Tarragona
\and Departament de Fonaments d'An\`alisi Econ\`omica, Universitat d'Alacant, Alacant}

\maketitle              


\section{Introduction}\label{sec_intro}

In recent years, the study of static and dynamical properties of complex networks
has received a lot of attention \cite{watts98,barabasi99,amaral00,albert02,dorogovtsev02}.
Complex networks appear in such diverse disciplines as sociology,
biology, chemistry, physics or computer science. In particular, great effort has been exerted
to understand the behavior of technologically based communication networks such as the Internet \cite{faloutsos99},
the World Wide Web \cite{albert99}, or e-mail networks \cite{newman02,ebel02,guimera??b}.
However, the study of communication processes in a wider sense is also of interest in other fields, remarkably the design of
organizations \cite{radner93,decanio98}. For instance, it is estimated that more than a half of the U.S. work force is dedicated to
information processing, rather than to make or sell things in the narrow sense \cite{radner93}.

The pioneering work of Watts and Strogatz \cite{watts98} opened a completely new field of research.
Its main contribution was to show that many real-world networks have properties of random graphs and
properties of regular low dimensional lattices. A model that could explain this observed behavior was
missing and the proposed "small-world" model of the authors
turned the interest
of a large number of scientist in the statistical mechanics
community in the direction of this appealing subject. Nevertheless, this simplified model
gives rise to a connectivity distribution function with an exponential form,
whereas many real world networks show a highly skewed degree distribution, usually with a power law tail
\begin{equation}
P(k)\propto k^{-\gamma}
\end{equation}
with an exponent $2\leq\gamma\leq 3$.
Barabasi and Albert \cite{barabasi99} proposed a model where nodes and links are added to the network in such a way that
the probability of the added nodes to be linked to the old nodes depend on the number of existing connections
of the old node. This simple computational model can explain the power law with an exponent $\gamma= 3$.

Tools taken from statistical mechanics have been used to understand not only the topological properties of these
communication networks, but also their dynamical properties. The main focus has been in the problem of
searchability, although when the number of search problems that the network is trying to solve increases it
raises the problem of congestion at some central nodes.
It has been observed, both in real world networks \cite{jacobson88} and in model communication networks
\cite{ohira98,tretyakov98,arenas01,sole01,guimera02},
that
the networks collapse when the load is above a certain threshold and the observed transition can be related to the appearance of
the $1/f$ spectrum of the fluctuations in Internet flow data \cite{csabai94,takayasu96}.

These two problems, search and congestion, that have so far been analyzed separately in the literature can be incorporated in the
same communication model. In previous works \cite{arenas01,guimera01,guimera02,guimera02b}
we have introduced a collection of models that captures the essential features of
communication processes and are able to handle these two important issues simultaneously.
In these models, agents are nodes of a network and can interchange information packets along the network links.
Each agent has a certain
capability that decreases as the number of packets to deliver increases. The transition from a free phase to a congested phase
has been studied for different network architectures in \cite{arenas01,guimera02}, whereas in \cite{guimera01} the cost of maintaining
communication channels was considered. Finally in \cite{guimera02b} we have attacked the problem of network
optimization for fixed number of links and nodes.

This paper is organized as follows. In Sect. 2 we present well known results about search in complex networks, whereas in Sect. 3
we review recent work on network load, being considered as a betweenness centrality and hence a
static characterization of the network. We present the common trends of our communication model in Sect. 4. In the next section,
we show some of the exact results that have been obtained for a particular class of network, Cayley trees. Finally, in the
last two sections we focus on the problem of network optimization, in the first one through a parameterized set of networks,
including connectivities that can be short- or long-ranged, and different degrees of preferentiallity, and
in the second one we perform an exhaustive search of optimal networks for a fixed number of nodes and links.

\section{Search in complex networks}\label{sec_search}

After the {\it discovery} of complex networks, one of the issues that
has attracted a lot of attention is ``search''. Real complex
communication networks such as the Internet or the World Wide Web are
continuously changing and it is not possible to draw a {\it map}
that allows to navigate in them. Rather, it is necessary to develop
algorithms that efficiently search for the desired computers or the
desired contents.

The origin of the study of this problem is in sociology since
the seminal experiment of Travers and Milgram \cite{travers69}.
Surprisingly, it was found that the average
length of acquaintance chains was about six. This means
not only that short chains exist in social networks as reported, for
example, in the ``small world'' paper by Watts and Strogatz
\cite{watts98}, but even more striking that these short chains can be
found using local strategies, that is without knowing exactly the
whole structure of the social network.

\begin{figure}[t]
\centerline{\includegraphics*[width=0.3\columnwidth]{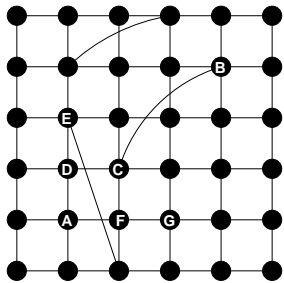}}
\caption{Network topology and search in Kleinberg's scenario. Consider
nodes $A$ and $B$. The distance between them is $\Delta_{AB}=6$
although the shortest path is only 3. A search process to get from $A$
to $B$ would proceed as follows. From $A$, we would jump with equal
probability to $D$ or $F$, since $\Delta_{DB}=\Delta_{FB}=5$: suppose
we choose $F$. The next jump would then be to $G$ or $C$ with equal
probability since $\Delta_{CB}=\Delta_{GB}=4$, although from $C$ it is
possible to jump directly to $B$. This is a consequence of the local
knowledge of the network assumed by Kleinberg.}
\label{kleinberg}
\end{figure}

The first attempt to understand theoretically the problem of {\it
searchability} in complex networks was provided by Kleinberg
\cite{kleinberg00}. In his work, Kleinberg proposes a
scenario where the network is modeled as a combination of a
two-dimensional regular lattice plus a number of long-range links. The
distance $\Delta_{ij}$ between two nodes $i$ and $j$ is defined as the
number of ``lattice-steps'' separating them in the regular lattice,
that is disregarding long-range links (see Fig. \ref{kleinberg}).
Long range links are not established at random. Instead, when a node
$i$ establishes one of such links, it connects with higher probability
with those nodes that are closer in terms of the distance $\Delta$. In
particular, the probability that the link is established with node $j$
is
\begin{equation}
\Pi_{ij}\propto\left(\Delta_{ij}\right)^{-r}
\end{equation}
where $r$ is a parameter.

The search algorithm proposed by Kleinberg is the following. A packet
standing at one node will be sent to the neighbor of the node that is
closer to the destination in terms of the distance $\Delta$. The
algorithm is local because, as shown in Fig. \ref{kleinberg}, the
heuristics of minimizing $\Delta$ does not warrant that the packet
will follow the shortest path between its current position and its
destination. Therefore, the underlying two-dimensional lattice has an
imprecise global informational content.

Kleinberg showed that with this essentially local scenario (with
imprecise global information), short paths cannot be found in general,
unless the parameter $r$ is fixed to $r=2$. This raised the
question of why real networks are then searchable, that is, how is it
possible that in real networks local strategies are able to find paths
that scale as $\log N$, where $N$ is the size of the
network. Recently, Watts and coworkers have shown that with an idea similar
to Kleinberg's, one can easily obtain searchable networks
\cite{watts02}. Their contribution consists in substituting the
underlying low-dimensional lattice by an {\it ultra-metric} space where
individuals are organized in a hierarchical fashion according to their
preferences, similitudes, etc. In this case, a broad collection of
networks turn out to be searchable.

Parallel to these efforts, there have been some attempts to exploit
the scale free nature of some networks to design algorithms that,
being local in nature, are still quite efficient
\cite{adamic01,tadic01}. The idea in all these works is to
profit from the scale-free nature of networks such as the Internet
and bias the search towards those nodes that have a high connectivity
and therefore act as hubs.

\section{Load and congestion in complex networks}

When the network has to tackle several simultaneous (or
parallel) search problems it raises the important issue of
congestion at overburdened
nodes \cite{jacobson88,ohira98,tretyakov98,arenas01,sole01}.
Indeed, for a single search problem the optimal network is
clearly a highly centralized star-like structure, with one or various
nodes in the center and all the rest connected to them. This structure
is {\it cheap} to assemble in terms of number of links and efficient
in terms of searchability, since the average cost (number of steps) to
find a given node is always bounded (2 steps), independently of the
size of the system. However, the star-like structure will
become inefficient when many search processes coexist in parallel in
the network, due to the limitation of the central node to process all the
information.

Load, independently of search, has been analyzed in different classes of networks
\cite{newman01,goh01,szabo02,goh02}.
The load, as introduced in these works, is equivalent to the betweenness
as it has been defined in social networks \cite{freeman77,newman01}.
The betweenness of a node $j$, $\beta_j$, is defined as
the number of minimum paths connecting pairs of nodes in the network
that go through node $j$.
Among the topological properties of networks, betweenness has become one of their main characteristics. In principle the
time needed for the computation of the betweenness of all vertices is of order $\mathcal{O}(MN^2)$, where $N$ is the number of nodes
and $M$ the number of links of the network. However, Newman \cite{newman01} introduced an algorithm
that reduces the magnitude of the time needed for the computation by a factor of $N$. This definition was used to measure the social
role played by scientists in some collaboration networks \cite{newman01}. Later on, it was also applied to quantify model networks.
Thus, in \cite{goh01} different networks are constructed and their distribution of betweennesses (or loads) measured. For instance,
scale-free networks with an exponent $2<\gamma \leq 3$ lead to a load distribution which is also a power law,
$P(\ell)\sim \ell ^{-\delta}$ with $\delta \approx 2.2$.
On the other side, the load distribution of small-world networks shows
a combined behavior of two Poisson-type decays. In subsequent work, the authors in \cite{goh02} suggested that real-world networks
should be classified in two different universality classes, according to the exponent of the power-law distribution of loads.
Finally, the distribution of loads was analytically computed for scale-free trees in \cite{szabo02}.

The works discussed in the previous paragraph consider the betweenness as a topological property of the network, since
it accounts for the number of shorter-paths going through a node.
However, to take into account the search algorithm and the fact that packets can perform several
random steps and then go through the same node more than once we introduce
an \emph{effective} betweenness.
The \emph{effective}  betweenness of node $j$, $B_j$,
represents the total number of packets that would pass through $j$ if
one packet would be generated at each node at each time step with
destination to any other node.
The  \emph{effective}  betweenness coincides with the \emph{topological} betweenness when the nodes have complete information of the
network structure and packets always  follow the shortest paths between origin and destination.

\section{A model of communication}\label{sec_model}

The model that can handle search and congestion at the same time
%
considers that the information is formed by discrete
packets that are sent from an origin node to a destination node. Each
node can store as many information packets as needed. However, the
capacity of nodes to deliver information cannot be infinite. In other
words, any realistic model of communication must consider that
delivering, for instance, two information packets takes more time than
delivering just one packet. A particular example of this would be to
assume that nodes are able to deliver one (or any constant number)
information packet per time step independently of their load, as
happens in the communication model by Radner \cite{radner93} and in
simple models of computer queues \cite{ohira98,tretyakov98,sole01},
but note that many alternative situations are possible. In the present
model, each node has a certain {\it capability} that decreases as the
load of accumulated packets increases. This limitation in the
capability of agents to deliver information can result in congestion
of the network. Indeed, when the amount of information is too large,
agents are not able to handle all the packets and some of them remain
undelivered for extremely long periods of time. The maximum amount of
information that a network can manage gives a measure of the quality
of its organizational structure. In the study of the model, the
interest is focused in both {\em when} the congestion occurs and {\em
how} it occurs.

\subsection{Description of the model}\label{subsec_model}

%

The dynamics of the model is as follows. At each time step $t$, an
information packet is created at every node with probability
$\rho$. Therefore $\rho$ is the control parameter: small values of
$\rho$ correspond to low density of packets and high values of $\rho$
correspond to high density of packets. When a new packet is created, a
destination node, different from the origin, is chosen randomly in the
network. Thus, during the following time steps $t+1,\,t+2,\ldots
,\,t+T$, the packet {\it travels} toward its destination. Once the
packet reaches the destination node, it is delivered and disappears
from the network. Another interpretation is possible for this
information transfer scenario. Packets can be regarded as problems
that arise at a certain ratio anywhere in an organization. When one of
such problems arises, it must be solved by an arbitrary agent of the
network. Thus, in subsequent time steps the problem flows toward its
{\it solution} until it is actually solved. This problem solving
scenario can be considered a particularly illustrative case of the more
general information transfer scenario. The problem solving
interpretation suggest a model similar to Garicano's \cite{garicano00}
in that there is task diversity and agents are specialized in solving
only certain types of tasks.

The time that a packet remains in the network is related not only to
the distance between the source and the target nodes, but also to the
amount of packets in its path. Indeed, nodes with high
loads---i.e. high quantities of accumulated packets---will need long
times to deliver the packets or, in other words, it will take long
times for packets to cross regions of the network that are highly
congested. In particular, at each time step, all the packets move from
their current position, $i$, to the next node in their path, $j$, with
a probability $q_{ij}$. This probability $q_{ij}$ is called the {\it
quality of the channel} between $i$ and $j$, and is defined as
\begin{equation}
q_{ij}=\sqrt{k_{i}k_{j}}\,,
\label{quality}
\end{equation}
where $k_{i}$ represents the {\it capability} of agent $i$ and, in
general, changes with time. The quality of a channel is, thus, the
geometric average of the capabilities of the two nodes involved, so
that when one of the agents has capability 0, the channel is
disabled.
It is assumed that $k_{i}$ depends only on the number of packets at
node $i$, $\nu_i$, through:
\begin{equation}
k_{i}=f(\nu_i)
\label{capability}
\end{equation}
The function $f(n)$ determines how the capability evolves when the
number of packets at a given node changes. In \cite{guimera02} we proposed a general form
although in this paper we will only show results for the case in which
the number of delivered packets is constant.
This particular case is consistent with simple
models of computer queues \cite{ohira98}, although the precise
definition of the models may differ from ours.

The election of the functional form for the quality of the channels
and the capability of the nodes is arbitrary. Regarding the first,
(\ref{quality}) is plausible for situations in which an
effort is needed from both agents involved in the communication
process. If, on the contrary,
information can be transmitted without the collaboration of the
receiver, an equation of the form
\begin{equation}
q_{ij}=k_{i}\,,
\label{quality2}
\end{equation}
would be more adequate. Equation (\ref{quality2}) will be used for
analytical understanding of the problem in Sect. \ref{sec_optim},
whereas (\ref{quality}) is used in Sect. \ref{sec_cayley}. Some of the
most relevant features of the model, however, are not dependent on which
one is used.

\subsection{Congestion and network capacity}

Depending on the ratio of generation
of packets $\rho$, two different behaviors are observed. When
the amount of packets is small, the
network is able to deliver all the packets that are generated and,
after a transient, the total load $N$ of the network achieves a
stationary state and fluctuates around a constant value. These
fluctuations are indeed quite small. Conversely, when $\rho$ is large
enough the number of generated packets is larger than the number of
packets that the network can manage to solve and the network enters a
state of congestion. Therefore, $N$
never reaches the stationary state but grows indefinitely in time. The
transition from the {\it free regime}, $\rho$ small, to the {\it
congested regime}, $\rho$ large, occurs for a well defined value of
$\rho$, that will be denoted $\rho_c$. For values smaller than but
close to $\rho_c$, the steady state is reached but large fluctuations
arise.

\begin{figure}[tb]
\centerline{\includegraphics*[width=.75\columnwidth]{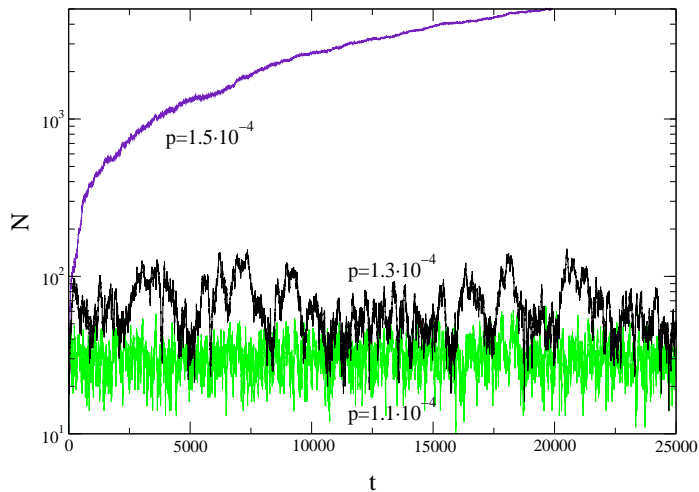}}
\caption{Evolution of the total number of packets, $N$, as a function
of time for a (5,7) Cayley tree and different values of $\rho$, below
the critical congestion point ($\rho=1.1\cdot 10^{-4}<\rho_c$), above
the critical congestion point ($\rho=1.5\cdot 10^{-4}>\rho_c$), and
close to the critical congestion point ($\rho=1.3\cdot 10^{-4}\approx
\rho_c$). Note the logarithmic scale in the $Y$ axis.}
\label{N_t}
\end{figure}

The three behaviors (free, congested and close to the transition) are
depicted in Fig. \ref{N_t}. For $\rho<\rho_c$, the width of the
fluctuations is small, indicating short characteristic times. This
means, among other thinks, that the average time required to deliver a
packet to the destination is small. It also means that
correlation times are short, that is, the state of the network at one time
step has little influence on the state of the network only a few time
steps latter. As $\rho$ approaches $\rho_c$, the fluctuations are
wider and one can conclude that correlations become important. In
other words, as one approaches $\rho_c$ the time needed to deliver a
packet grows and the state of the network at one instant is
determinant for its state many time steps later. In the congested
regime, the amount of delivered packets is independent of the load
and thus remains constant over time, while the number of generated
packets is also constant, but larger than the amount of delivered
packets. Thus, at each time step the number of accumulated packets is
increased by a constant amount, and $N(t)$ grows linearly in time.

The transition from the free regime to the congested regime is
therefore captured by the slope of $N(t)$ in the stationary
state. When all the packets are delivered and there is no
accumulation, the average slope is 0 while it is larger than 0 for
$\rho>\rho_c$. We use this property to introduce an {\it order
parameter}, $\eta$, that is able to characterize the transition from
one regime to the other:
\begin{equation}
\eta(p)=\lim_{t\rightarrow\infty}\frac{1}{\rho S}\frac{\left\langle\Delta N\right\rangle}{\Delta t},
\label{def_order}
\end{equation}
In this equation $\Delta N=N(t+\Delta t)-N(t)$, $\langle\ldots\rangle$
indicates an average over time windows of width $\Delta t$ and $S$ is the
number of nodes in the system. Essentially, the order
parameter represents the ratio between undelivered and generated
packets calculated at long enough times such that $\Delta
N\propto\Delta t$. Thus, $\eta$ is only a function of the probability
of packet generation per node and time step, $\rho$. For
$\rho>\rho_c$, the system collapses, $\langle\Delta N\rangle$ grows
linearly with $\Delta t$ and thus $\eta$ is a function of $\rho$
only. For $\rho<\rho_c$, $\langle\Delta N\rangle=0$ and $\eta=0$.
Since the order parameter is continuous at $\rho_c$,
the transition to congestion is a critical phenomenon and $\rho_c$ is a
critical point as usually defined in statistical mechanics
\cite{stanley87}.

Once the transition is characterized, the first issue that deserves
attention is the location of the transition point $\rho_c$ as a
function of the parameters of the network. This
transition point gives information about the capacity of a given
network. Indeed, the maximum number of packets that a network can
handle per time step will be $N_c=S\rho_c$. Therefore, $\rho_c$ is a
measure of the amount of information an organization is able to handle
and thus of the efficiency of a given organizational structure. One
reasonable problem to propose is, therefore, which is the network that
maximizes $\rho_c$ for a fixed set of available resources (agents
and links).

\section{Analytical results for hierarchical lattices}\label{sec_cayley}

\begin{figure}[tb]
\centerline{\includegraphics*[width=.75\columnwidth]{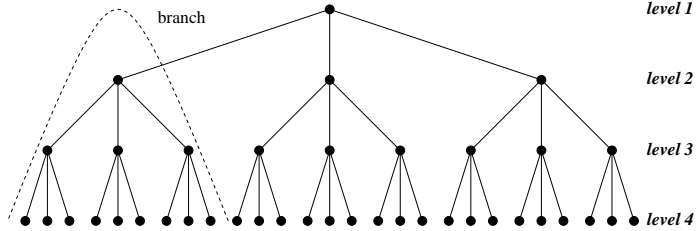}}
\caption{Typical hierarchical tree structure used for simulations and
calculations: in particular, it is a tree $(3,4)$. Dashed line:
definition of branch, as used in some of the calculations.}
\label{cayley}
\end{figure}

As a first step we considered hierarchical networks, since they provide a zeroth order approximation
to real structures, and have also been used in the economics literature to model organizations \cite{radner93,bolton94}.
In particular we are going to focus on hierarchical Cayley trees, as depicted in Fig. \ref{cayley}.
Cayley trees are identified by their branching $z$ and their number of levels $m$,
and will be denoted $(z,m)$ hereafter.

In this case the system is regarded as hierarchical also from a knowledge point of view. It
is assumed in the model that agents have complete knowledge of the
structure of the network in the subbranch they root. Therefore, when
an agent receives a packet, he or she can evaluate whether the
destination is to be found somewhere below. If so, the packet is sent
in the right direction; otherwise, the agent sends the packet to his
or her supervisor. Using this simple routing algorithm, the packets
travel always following the shortest path between their origin and
their destination.

\begin{figure}[tb]
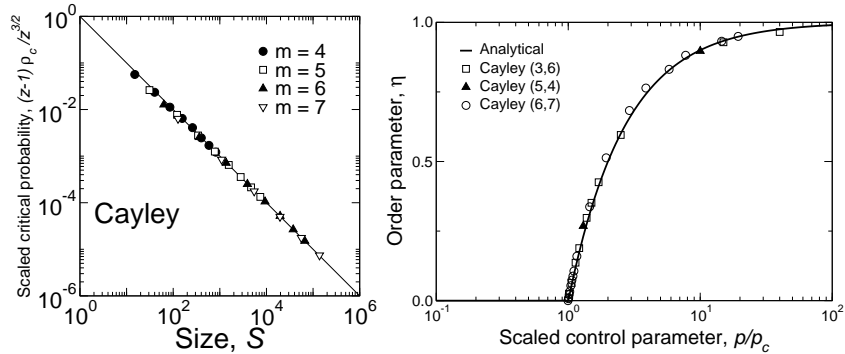

\centerline{\includegraphics*[width=0.4\columnwidth]{pc_bethe}
\includegraphics*[width=0.50\columnwidth]{par_ordre}}
\caption{Comparison between analytical (lines) and numerical (symbols)
values obtained for hierarchical trees. Left: scaled critical probability (\ref{p_c_value}).
Right: order parameter (\ref{ord_eq}).
\label{par_ordre}}
\end{figure}

As happens in other problems in statistical physics \cite{stauffer92},
the particular symmetry of the hierarchical tree allows an analytical
estimation of the critical point $\rho_c$. In particular, the approach
taken here is {\it mean field} in the sense that fluctuations are
disregarded and only average expected values are considered.
By using the steady state condition that the number of packets arriving at the top node, which is the
most congested one, equals the number of packets leaving it we arrive to the following inequality
\begin{equation}
\rho_c\geq\frac{\sqrt{z}}{\frac{z(z^{m-1}-1)^2}{z^m-1}+1}
\label{p_c_value}
\end{equation}
when the quality of the channels is given by (\ref{quality}).
Although this expression  provides an upper
bound to $\rho_c$, (\ref{p_c_value}) is an excellent
approximation for $z\ge 3$, as shown  in Fig. \ref{par_ordre}.

The total critical number of generated packets, $N_c=\rho_cS$, with $S$
denoting the size of the system, can be approximated, for large
enough values of $z$ and $m$ such that $z^{m-1}\gg1$, by
\begin{equation}
N_c=\frac{z^{3/2}}{z-1},
\label{N_c}
\end{equation}
which is independent of the number of levels in the tree. It suggests
that the behavior of the top node is only affected by the total number
of packets arriving from each node of the second level, which is
consistent with the mean field hypothesis.

According to (\ref{N_c}), the total number of packets a
network can deal with, $N_c$, is a monotonically increasing function
of $z$, suggesting that, given the number of agents in the
organization, $S$, the optimal organizational structure, understood as
the structure with highest capacity to handle information, is the
flattest one, with $m=2$ and $z=S-1$.

To understand this result it is necessary to take into account the
following considerations:
\begin{itemize}
\item We are restricting our comparison only to different hierarchical
networks and in any hierarchical network, the top node will receive
most of the packets. Since origins and destinations are generated with
uniform independent probabilities, roughly $(z-1)/z$ of the packets
will pass through the top node.
\item Still, it could seem that having small $z$ is {\it slightly}
better according to the previous consideration. However, it is
important to note that, in the present model (in particular due to
(\ref{quality})), the loads of both the sender and the
receiver are important to have a good communication quality. In a
network with small $z$, the nodes in the second level have also a high
load, while in a network with a high $z$ the nodes in the second level
are much less loaded.
\item We have implicitly assumed that there is no cost for an
agent to have a large amount of communication channels active.
\end{itemize}

For the order parameter, it is possible to derive an
analytical expression for the simplest case where there are only two
nodes that exchange packets. Since from symmetry considerations
$\nu_1=\nu_2$, the average number of packets eliminated in one time step
is $2$, while the number of generated packets is $2\rho$. Thus $\rho_c=1$
and with the present formulation of the model it is not possible to
reach the super-critical congested regime. However, $\rho$ can be
extended to be the average number of generated packets per node at
each step (instead of a probability) and in this case it can actually
be as large as needed. As a result, the order parameter for the
super-critical phase is $\eta=(\rho-1)/\rho$. As observed in
Fig.~\ref{par_ordre}, the general form
\begin{equation}
\eta(\rho/\rho_c)=\frac{\rho/\rho_c -1}{\rho/\rho_c}
\label{ord_eq}
\end{equation}
fits very accurately the behavior of the order parameter for any
Cayley tree.

\section{Optimization in model networks}\label{wehia}

In this section we extend previous studies about local search in model
networks in two directions. First, we consider networks that, as in
Kleinberg's work, are embedded in a two-dimensional space, but study
the effect not only of long range random links but also of long range
preferential links. Secondly and more significantly, we
consider the effect of congestion when multiple searches are carried
out simultaneously. As we will show, this effect has drastic
consequences for optimal network design.

\subsection{Network topology}\label{nettop}

The small world model \cite{watts98} considered
two main components: local linking with neighbors and random long
range links giving rise to short average distance between nodes. The
idea of Kleinberg is that local linking provides information about the
social structure and can be exploited to heuristically direct the
search process. Later, Barabasi and Albert showed that growth and
preferential attachment play a fundamental role in the formation of
many real networks \cite{barabasi99}. Even though this model captures
the correct mechanism for the emergence of highly-connected nodes, it
is not likely that it captures all mechanisms responsible for the
evolution of ``real-world'' scale-free networks. In particular, it
seems plausible that in many of the networks that show scale-free
behavior there is also an underlying structure as in the Watts and Strogatz model. To
illustrate this idea, consider web-pages in the World Wide Web.  It is
plausible to assume that a page devoted to physics is more likely to
be connected to another page devoted to physics than to a page devoted
to sociology. That is, a set of pages devoted to physics is likely to be
more inter-connected than a set including pages devoted to physics and
sociology.

\begin{figure}[t]
\centerline{\includegraphics*[width=0.8\columnwidth]{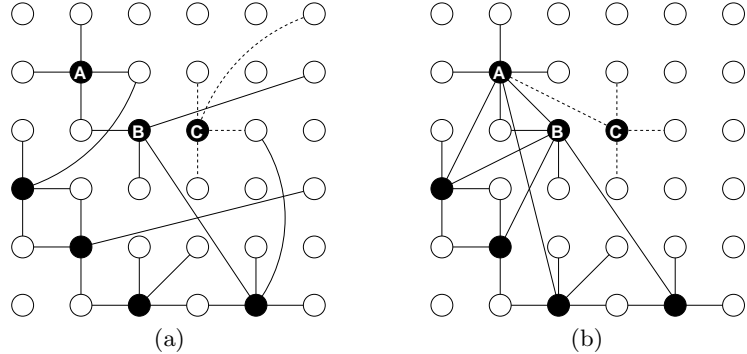}}
\centerline{(a)\hspace{0.42\columnwidth}(b)}
\caption{Construction of networks with multiple linking mechanisms. In
both cases $\phi=0.25$. A random node is selected at each time step
and $m=4$ new links starting from that node are created. Black nodes
represent nodes that have already been selected. Dotted lines
represent the links created during the last time step in which node
$C$ was selected. In (a), the destination of long range links is
created at random ($\gamma=0$), while in (b) they are created
preferentially ($\gamma>0$) and nodes $A$ and $B$ are attracting most
of them.}
\label{creation}
\end{figure}

Therefore we consider networks with four basic components: growth,
preferential attachment, local attachment and random attachment. To
create the network the following algorithm is used:
\begin{enumerate}
\item Nodes are located in a two-dimensional square lattice without
interconnecting them.
\item A node $i$ is chosen at random.
\item We create $m$ links starting at the selected node. With
probability $\phi$, the destination node is selected
preferentially. With probability $1-\phi$ the destination node is one
of the nearest neighbors of the selected node. When the destination node is
selected preferentially, we apply the following rule: the probability
that a given destination node $j$ is chosen is a function
of its connectivity
\begin{equation}
\Pi_j\propto k_j^\gamma ,
\end{equation}
where $k_j$ is the number of links of node $j$ and $\gamma$ is a
parameter that allows to tune the network from maximum
preferentiallity to no preferentiallity. Indeed, for $\gamma=0$ the
links are random and for $\gamma=1$ we recover the BA model, that
generates scale free networks in the case $\phi=1$. For $\gamma>1$, a few
nodes tend to accumulate all the links.
\item A new node is chosen and the process is repeated from step 3,
until all the nodes have been chosen once.
\end{enumerate}
Figure \ref{creation} shows two examples of networks in the process of
being created according to this algorithm.
Note that in this case, the number of links is fixed and the existence
of long range links implies that some local links are not present and
therefore that the information contained in the two-dimensional
lattice is less precise.

\subsection{Communication model and search algorithm}

After the definition of the network creation algorithm, we move to the
specification of the communication model and the search algorithm. For
the communication model, we will use the general model presented and
discussed in Sect. \ref{sec_model}. As already stated, this model is
general enough and considers the effect of congestion due to limitation of ability of nodes
to handle information.

In comparison with hierarchical networks, there is only one ingredient
of the communication model that needs to be reformulated. In the
hierarchical version of the model, when a node receives a packet, it
decides to send it downwards in the right direction if the solution is
there, or upward to the agent overseeing her otherwise. This simple
{\it routing algorithm} arises from the fact that we implicitly assume
that the hierarchy is not only a communicational hierarchy, but also a
knowledge hierarchy, where nodes know perfectly the structure of the
network {\it below} them. In a complex network, this informational
content of the hierarchy is lost. Here we will use Kleinberg's
approach \cite{kleinberg00}. When an agent receives a
packet, she knows the coordinates in the underlying two-dimensional
space of its destination. Therefore, she forwards the packet to the
neighbor that is {\em closer} to the destination according to the
lattice distance $\Delta$ defined in Sect. \ref{sec_search},
provided that the packet has not visited that node
previously\footnote{Packets are sent to previously visited nodes only
if it is strictly necessary. This {\it memory} restriction avoids
packets getting trapped in loops}. Note, however, that distance
refers to the two-dimensional space, but not necessarily to the
topology of the complex network and, as in Kleinberg's work, there
might be shortcuts in directions that increase $\Delta$. Moreover,
here long range links {\it replace} short range links and are not
simply added to short range links. Therefore it is possible that
following the direction of minimization of $\Delta$ the packet arrives
to a dead end and has to go back.

Considering this algorithm, it is interesting that the three
mechanisms to establish links (local, random and preferential) are
somehow complementary. A completely regular lattice (all links are
local) contains a lot of information since all the agents efficiently
send their packets in the best possible direction. However, the
average path length is extremely high in this networks and therefore
the number of packets that are flowing in the network at a given time
is also very high. The addition of random links can reduce
dramatically the average path length, as in small
world networks. However, if the number of random links is very high, then the
number of local links is small and thus sending the packet to the node
closer to the destination is probably quite inefficient (since it is possible
that, even if it is very close in the underlying
two-dimensional space, there is no short path in the actual
topology of the network). Finally, preferential links seem to solve
both problems. They obviously solve the long average path length
problem but, in addition, the loss of information is not large,
since the highly connected that actually concentrate
this information. The star configuration is an extreme example of
this: although there are no local links, the central node is capable
of sending all the packets in the right directions. However, when the
amount of information to handle is big, preferential links are
especially inadequate because highly connected nodes act as centers of
congestion. Therefore, optimal structures should be networks where all
the mechanisms coexist: complex networks.

\subsection{Results}

We simulate the behavior of the communication model in networks built
according to the algorithm presented in Sect. \ref{subsec_model}. First, a
value of the probability of packet generation per node and time step,
$\rho$, is fixed. For that particular value, we compare the
performance of different networks: networks with different
preferentiallity, from random ($\gamma=0$) to maximum centralization
($\gamma\gg 1$), and with different fraction of long range links, from
pure regular lattices with no long range links ($\phi=0$) to networks
with no local component ($\phi=1$). For each collection of the
parameters $\rho$, $\gamma$, and $\phi$, the network load,
$\overline{N}$, is calculated and averaged over a certain time window
and over 100 realizations of the network, so that fluctuations due to
particular simulations of the packet generation and of the network
creation are minimized. As in the economics literature, the objective
is to minimize the average time $\tau$ for a packet to go from the origin to
the destination.

According to Little's Law of queuing theory \cite{allen90}, the
characteristic time is proportional to the average total load,
$\overline{N}$, of the network:
\begin{equation}
\frac{\overline{N}}{\tau}=\rho S\Rightarrow \tau=\frac{\overline{N}}{\rho S}
\end{equation}
where $\rho$ is the probability of packet generation for each node at each time
step. Thus, minimizing the average cost of a search is equivalent to
minimizing the total load $\overline{N}$ of the network.

\begin{figure}[t]
\centerline{\includegraphics*[width=0.45\columnwidth]{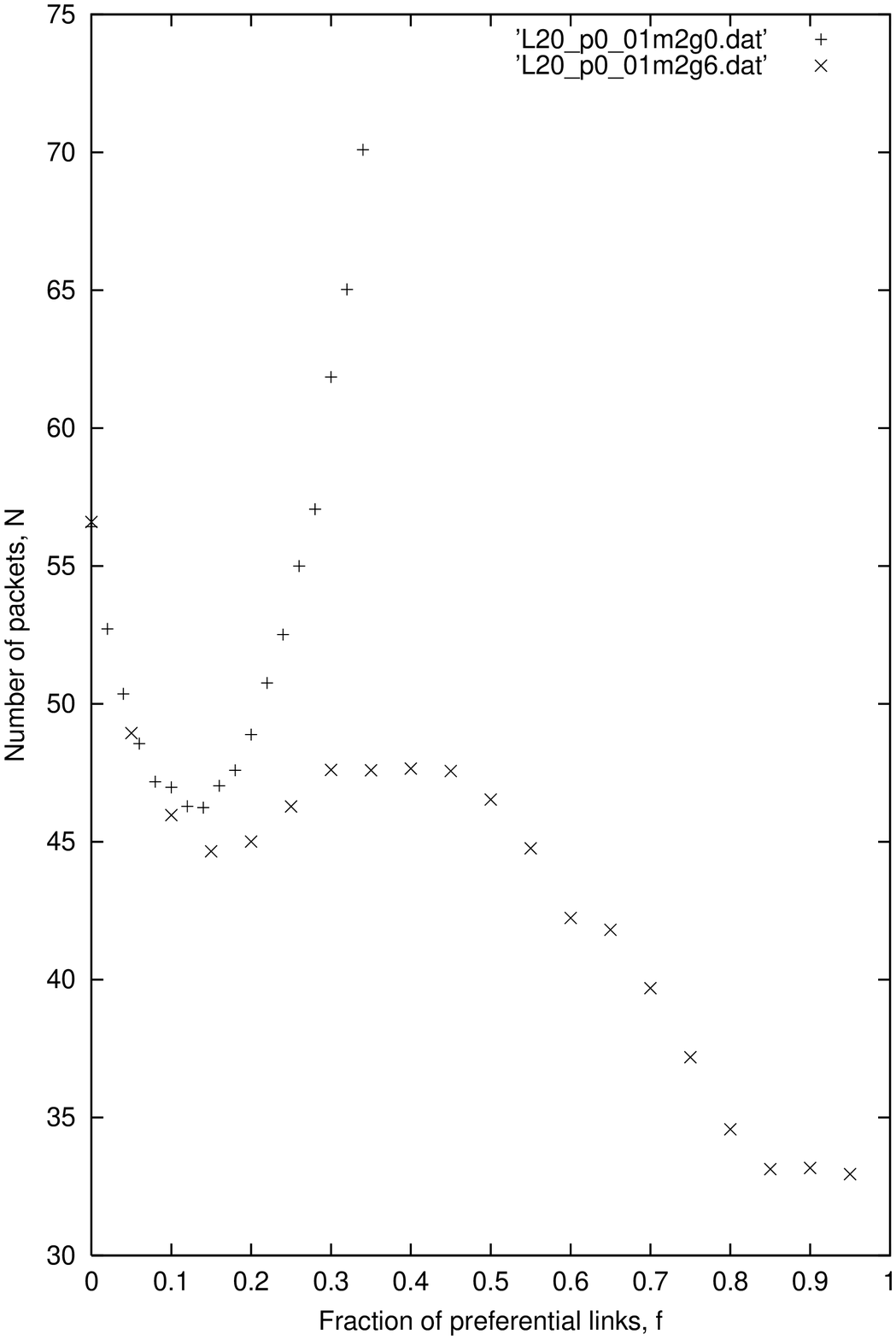}\quad\includegraphics*[width=0.45\columnwidth]{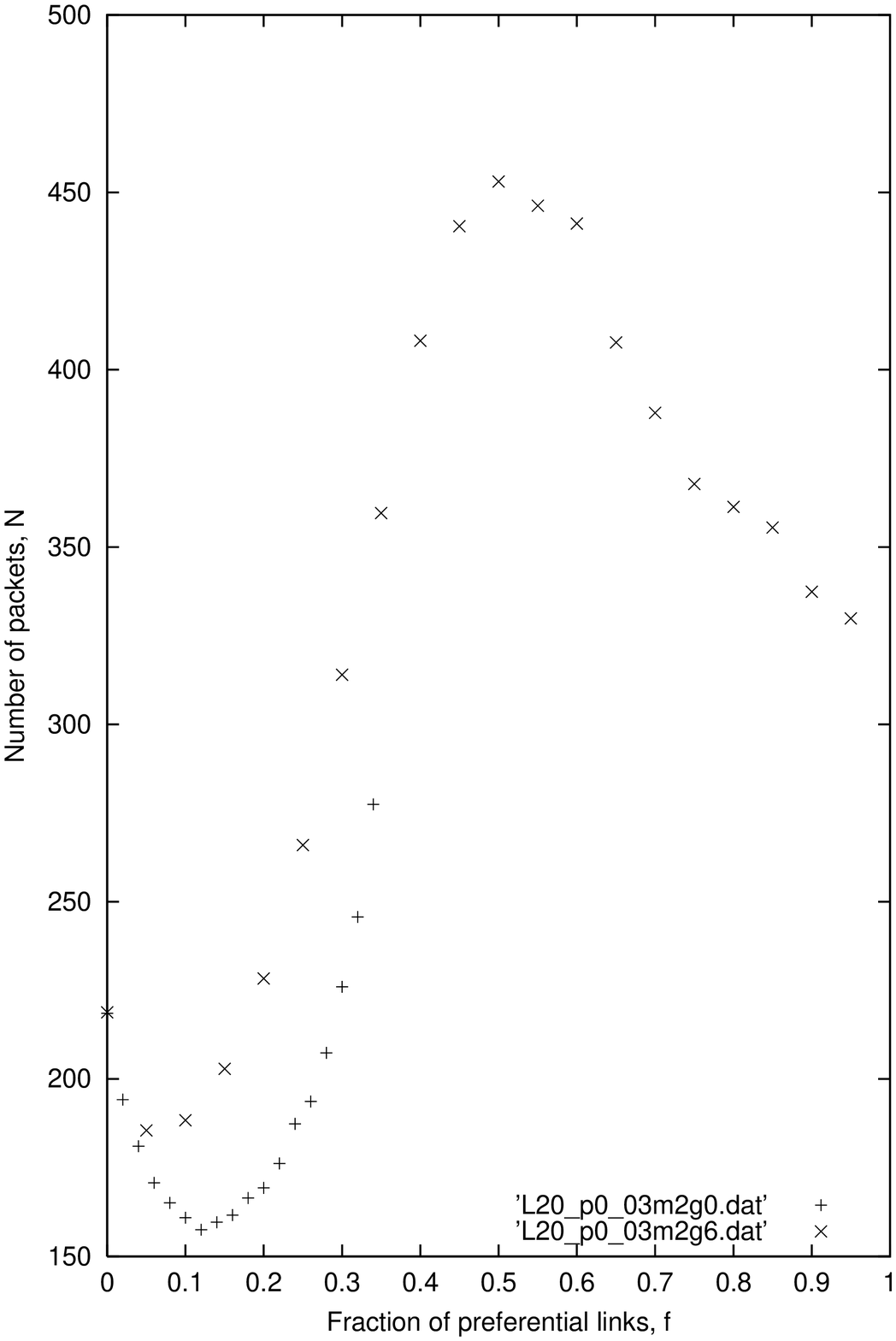}}
\centerline{(a)\quad\quad\quad\quad\quad\quad\quad\quad\quad\quad\quad\quad\quad(b)}
\centerline{\includegraphics*[width=0.32\columnwidth]{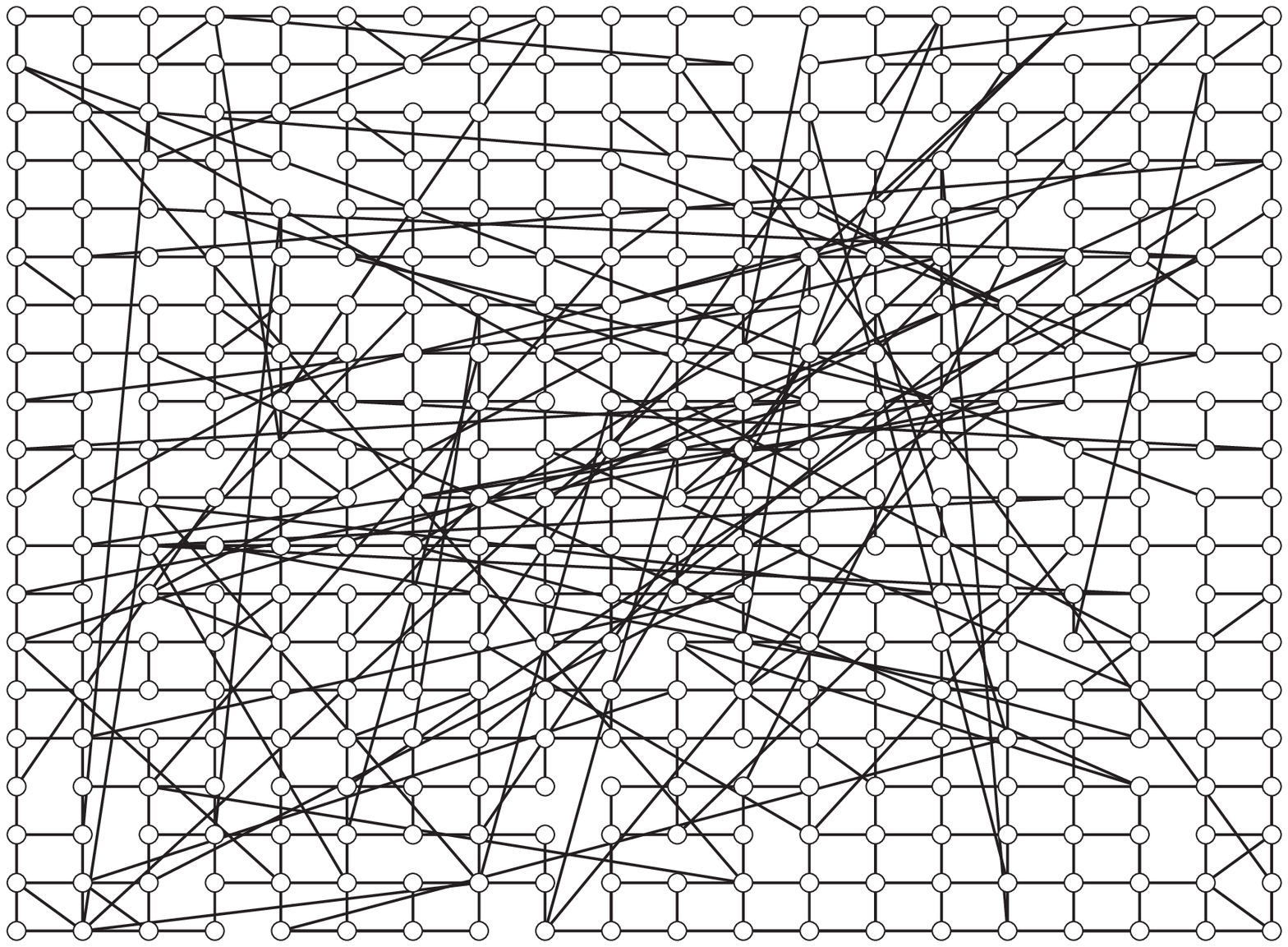}\includegraphics*[width=0.32\columnwidth]{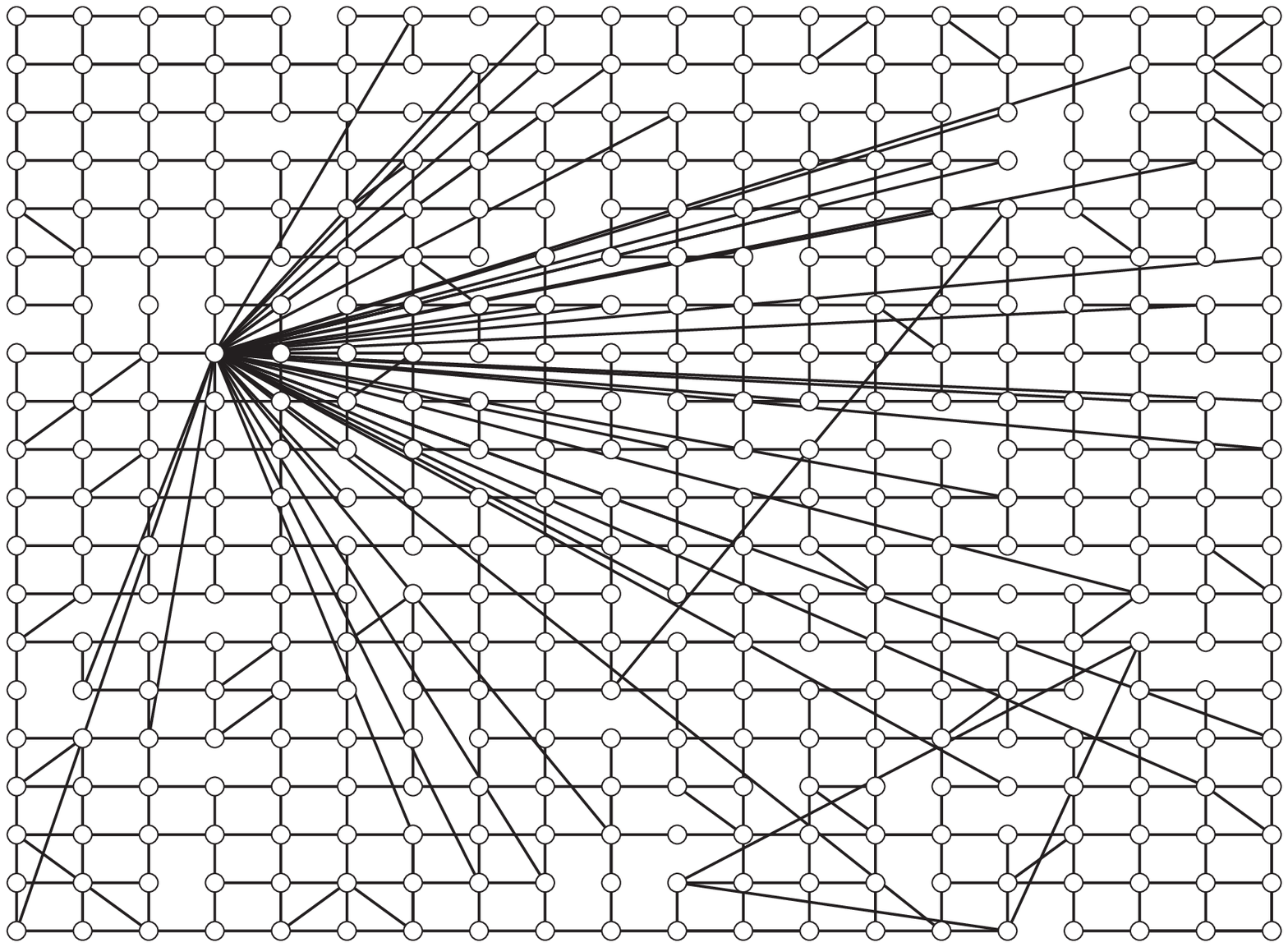}
\includegraphics*[width=0.32\columnwidth]{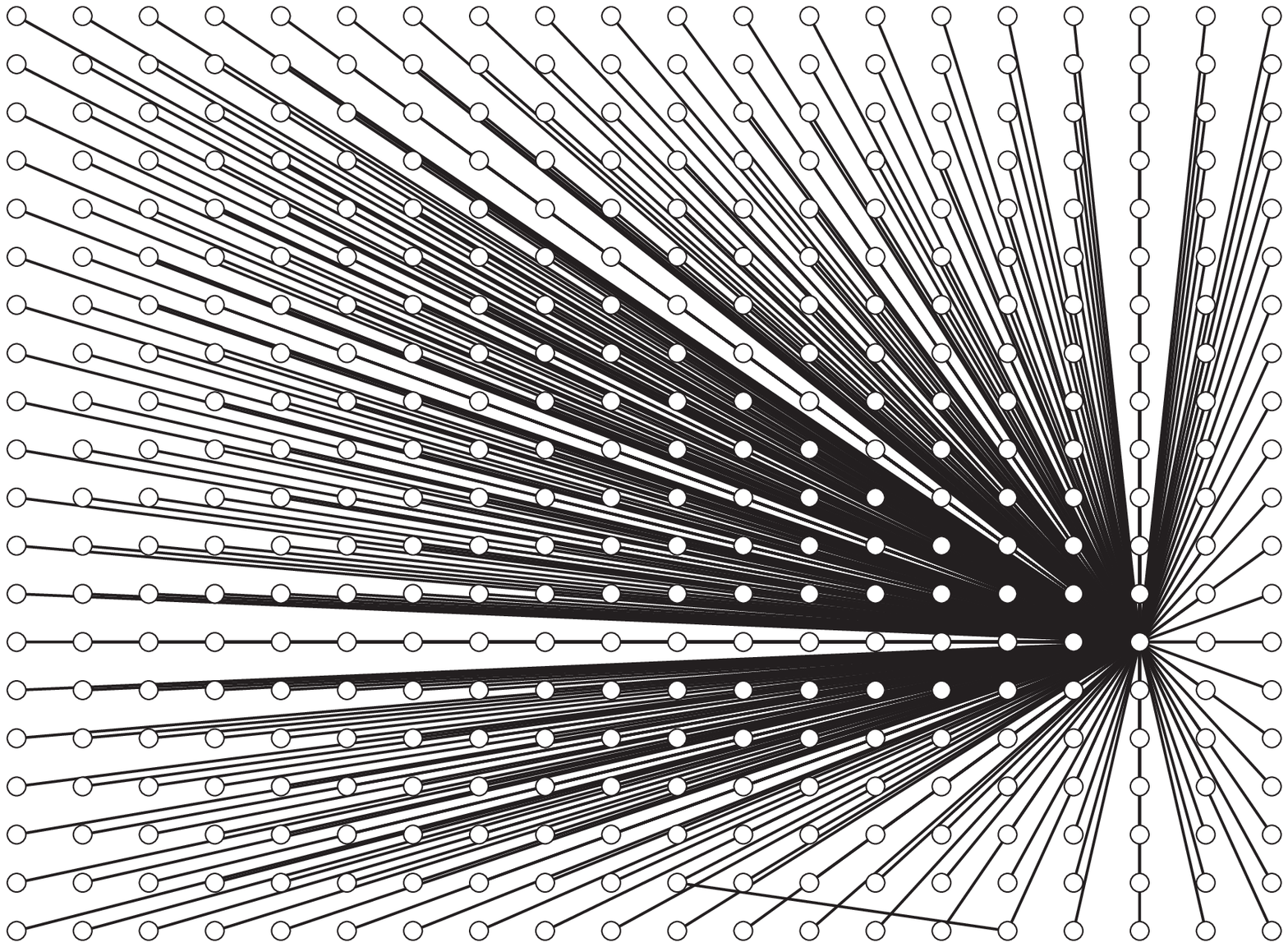}}
\centerline{(c)\quad\quad\quad\quad\quad\quad\quad\quad(d)\quad\quad\quad\quad\quad\quad\quad\quad(e)}
\caption{(a) and (b) Average number of packets flowing in the network
as a function of the fraction of preferential links: (a) $\rho=0.01$
and (b) $\rho=0.03$. Symbol ($+$) corresponds to $\gamma=0$ (random
links) and symbol ($\times$) corresponds to $\gamma=6$ (extremely
focused links). Figures (c),(d) and (e) show the typical shape of
complex networks with particularly efficient configurations: (c)
$\gamma=0$ and $\phi=0.12$; (d) $\gamma=6$ and $\phi=0.07$; and (e)
$\gamma=6$ and $\phi=1.0$;
\label{complex}}
\end{figure}

The main results are shown in Fig. \ref{complex}.
Consider first the behavior of the networks at low values of
$\rho$. Figure \ref{complex}.a shows the load of the network for
$\rho=0.01$ as a function of the fraction of long range links, $\phi$,
both when they are random $\gamma=0$ and when they are extremely
preferential $\gamma=6$. In the last case, long range links are
established only with the most connected node. In this case of small
$\rho$, centralization is not a big problem because congestion effects
are still not important. Therefore, preferential links are, in
general, better than random long range links. In the case of
preferential links, it is interesting to understand the behavior of
the curve $\overline{N}(\phi)$. For $\phi=0$ the network is a
two-dimensional regular lattice and then the average distance between
nodes is large. As some long range links are introduced, the average
path length decreases as in the Watts-Strogatz model \cite{watts98}, and
therefore the load of the network is smaller because packets reach
their destination faster. However, the addition of long range links
implies the lack of local links and when $\phi$ is further increased,
the heuristic of minimizing the lattice distance $\Delta$ becomes
worse and worse. This fact explains that for $\phi\approx 0.15$ (the
network is similar to the one depicted in Fig. \ref{complex}.d) the
load has a local minimum that arises due to the trade-off between the
two effects of introducing long range preferential links: shortening
of the distances that tends to decrease $\overline{N}$ and destruction
of the lattice structure that tends to decrease the utility of the
heuristic search and then to increase $\overline{N}$. If $\phi$ is
further increased, one node tends to concentrate all the links and for
$\phi=1$ (Fig. \ref{complex}.e) the network is strictly a star with
one central node and the rest connected to it. In this completely
centralized situation, the lack of two-dimensional lattice is not
important because the packets will be sent to the central node and
from there directly to the destination. Since for small $\rho$
congestion is not an issue, this structure turns out to be even better
than the locally optimal structure with $\phi\approx 0.15$.

The situation is different when considering higher values of the
probability of packet generation (Fig. \ref{complex}.b displays the
the results for $\rho=0.03$). Regarding preferential linking, the two
locally optimal structures with $\phi=0.7$ and $\phi=1$ (Figs.
\ref{complex}.d and \ref{complex}.e respectively) persist. However, in
this situation and due to congestion considerations the first is
better than the second. Thus, at some intermediate value of
$0.01<\rho<0.03$, there is a transition such that the optimal
structure changes from being the star configuration to being the {\it
mixed} configuration with local as well as preferential
links. Significantly, this transition is sharp, meaning that there is
not a continuous pass from the star to the mixed.

Beyond the behavior of networks built with preferential long range
links, it is worth noting that when the effect of the congestion is
important (Fig. \ref{complex}.b), the structure depicted in Fig.
\ref{complex}.c, where the long range links are actually thrown at
random, becomes better than the structure in \ref{complex}.d. In other
words, the optimal network is, in this case, a completely
decentralized small world network {\it a la} Watts-Strogatz.

\section{Optimization in a general framework}\label{sec_optim}

In the previous section we have compared the behavior of networks which
have been built following different rules (nearest neighbor linking, preferential attachment, etc.).
The main reason for focusing on a particular set of networks is that it is very costly to compare the
performance of two networks: it is necessary to run a simulation,
wait for the stationary state and calculate the average load of the network.
Specially, close to the critical point the time needed to reach the stationary state
diverges. In \cite{guimera02b} we presented a formalism that is able to
cope with search and congestion simultaneously, allowing the determination of
optimal topologies. This formalism avoids the problem of
simulating the dynamics of the communication process and provides a
general scenario applicable to any communication process.

Let us focus on a single information packet at node $i$ whose destination is node $k$.
The probability for the packet to go from $i$ to a new node $j$ in its next movement is $p_{ij}^k$.
In particular,
$p_{kj}^k=0$ $\forall j$ so that the packet is {\it removed} as soon
as it arrives to its destination. This formulation is completely
general, and the precise form of $p_{ij}^k$ will depend on the search
algorithm and on the connectivity matrix of the network.
In particular, when the search is Markovian, $p_{ij}^k$
does not depend on previous positions of the packet. In this case, the
probability of going from $i$ to $j$ in $n$ steps is given by
\begin{equation}
P_{ij}^k(n)=\sum_{l_1,l_2,\dots,l_{n-1}}p_{il_1}^k p_{l_1l_2}^k\cdots p_{l_{n-1}j}^k.
\end{equation}
This definition allows us to compute the average number of times, $b_{ij}^k$,
that a packet generated at $i$ and with destination at
$k$ passes through $j$.
\begin{equation}
b^k=\sum_{n=1}^\infty P^k(n)=\sum_{n=1}^\infty
\left(p^k\right)^n=(I-p^k)^{-1}p^k.
\label{bij}
\end{equation}
and the effective betweenness of node $j$, $B_j$, is then defined as the
sum over all possible origins and destinations of the packets,
\begin{equation}
B_j=\sum_{i,k}b_{ij}^k.
\label{B}
\end{equation}
When the search
algorithm is able to find the minimum paths between nodes, the
effective betweenness will coincide with the topological
betweenness, $\beta_j$, as usually defined
\cite{freeman77,newman01}.

Once, these quantities have been defined, we focus on the load of the network, $N(t)$, which is the number of
floating packets.
These floating packets are stored in the nodes that act as queues.
In a general scenario where packets are generated at random and independently at each node with
a probability $\rho$, the arrival of packets to a given node $j$ is a Poisson process.
In the original model presented in Sect. \ref{sec_model} we assumed that the quality of the channels
depend on both the sender and the receiver nodes; if one assumes that it only depends on the
receiver node then the delivery of packets is also a Poisson process. In this simple picture,
the queues are called M/M/1 in the computer science literature and the
average load of the network is \cite{allen90,guimera02b}
\begin{equation}
\overline{N}=\sum_{j=1}^S\frac{\frac{\rho B_j}{S-1}}{1-\frac{\rho B_j}{S-1}}.
\label{load}
\end{equation}
There are two interesting limiting cases of equation
(\ref{load}). When $\rho$ is very small,
taking into account that the sum of betweennesses is proportional to the average distance,
one obtains that the load is proportional to the average effective distance.
On the other hand, when $\rho$ approaches $\rho_c$ most of the load of
the network comes from the most congested node, and therefore
\begin{equation}
\overline{N}\approx\frac{1}{1-\frac{\rho B^*}{S-1}}\quad\quad\rho\rightarrow\rho_c,
\label{N2}
\end{equation}
where $B^*$ is the effective betweenness of the most central
node. The last results suggest the following interesting
problem: to minimize the load of a network it is necessary to minimize
the effective distance between nodes if the amount of packets is
small, but it is necessary to minimize the largest effective
betweenness of the network if the amount of packets is large. The
first is accomplished by a {\it star-like} network, that is, a network
with one central node and all the others connected to it. The
second, however, is accomplished by a very decentralized network in which all
the nodes support a similar load. This behavior is similar to any
system of queues provided that the communication depends only on the
sender.

It is worth noting that there are only two assumptions in the
calculations above. The first one has already been mentioned: the
movement of the packets needs to be Markovian to define the jump
probability matrices $p^k$. Although this is not strictly true in real
communication networks---where packets are not usually  allowed to go
through a given node more than once---it can be seen as a first
approximation \cite{ohira98,arenas01,sole01}. The second assumption is
that the jump probabilities $p_{ij}^k$ do not depend on the congestion
state of the network, although communication protocols sometimes try
to avoid congested regions, and then $B_j=B_j(\rho)$. However, all the
derivations above will still be true in a number of general
situations, including situations in which the paths that the packets
follow are unique, in which the routing tables are fixed, or
situations in which the structure of the network is very homogeneous
and thus the congestion of all the nodes is similar.
Compared to situations in which packets avoid congested
regions, it correspond to the worst case
scenario and thus provide bounds to more realistic scenarios in which
the search algorithm interactively avoids congestion.

Equation (\ref{load}) relates a dynamical variable, the load, with the topological
properties of the network and the properties of the algorithm.
So we have converted a dynamical communication problem into a topological problem.
Hence, the dynamical optimization procedure of finding the structure that gives the minimum load
is reduced to a topological optimization procedure where the network is characterized completely
by its effective betweenness distribution.
In \cite{guimera02b} we considered the problem of finding optimal structures for a purely local search,
using a generalized simulated
annealing (GSA) procedure, as described in \cite{tsallis94,penna95}.
On the one side, we have found that for
$\rho\rightarrow 0$ the optimal network has a star-like centralized
structure as expected, which corresponds to the minimization of the
average effective distance between nodes. On the other extreme, for
high values of $\rho$, the optimal structure has to minimize the
maximum betweenness of the network; this is
accomplished by creating a homogeneous network where all the nodes
have essentially the same degree, betweenness, etc.
One could expect that the transition centralized-decentralized occurs
progressively. Surprisingly, the results of the optimization process
reveal a completely different scenario. According to simulations, star-like configurations
are optimal for $\rho<\rho^*$; at this point, the homogeneous networks
that minimize $B^*$ become optimal. Therefore there are only two type
of structures that can be optimal for a local search process:
star-like networks for $\rho<\rho^*$ and homogeneous networks for
$\rho>\rho^*$.

\begin{figure}[t]
\centerline{\includegraphics*[width=0.25\columnwidth]{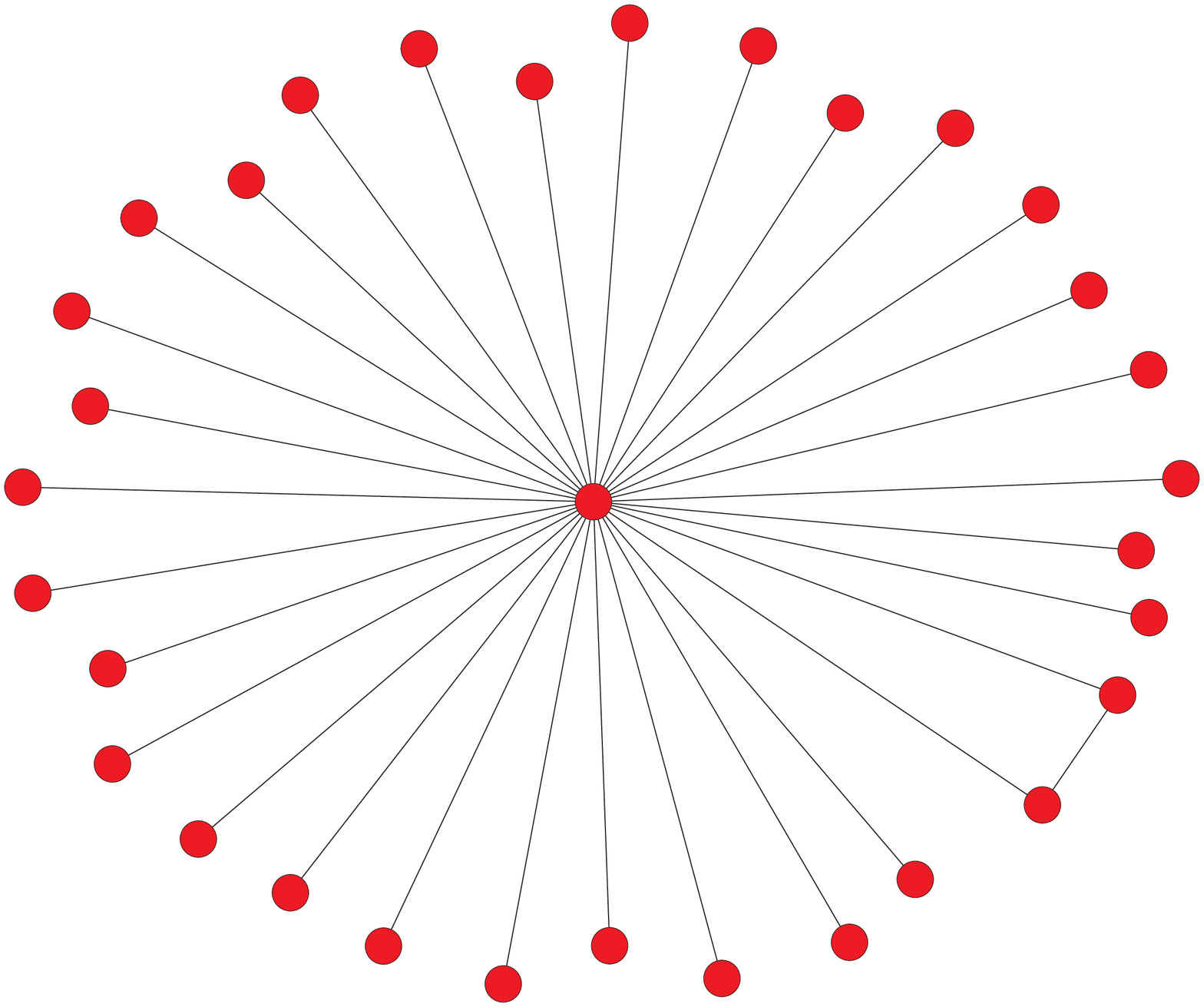}\includegraphics*[width=0.25\columnwidth]
{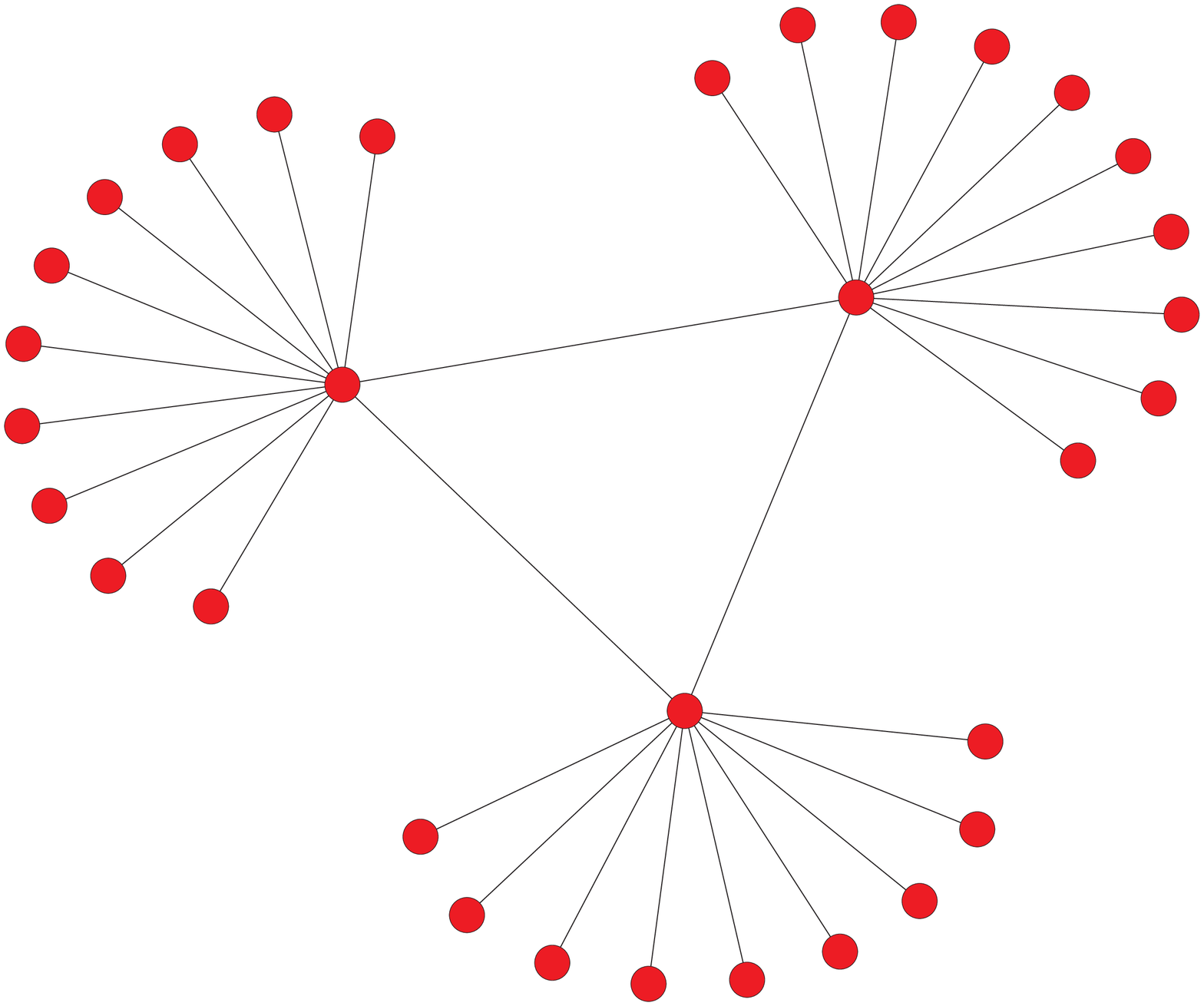}\includegraphics*[width=0.25\columnwidth]{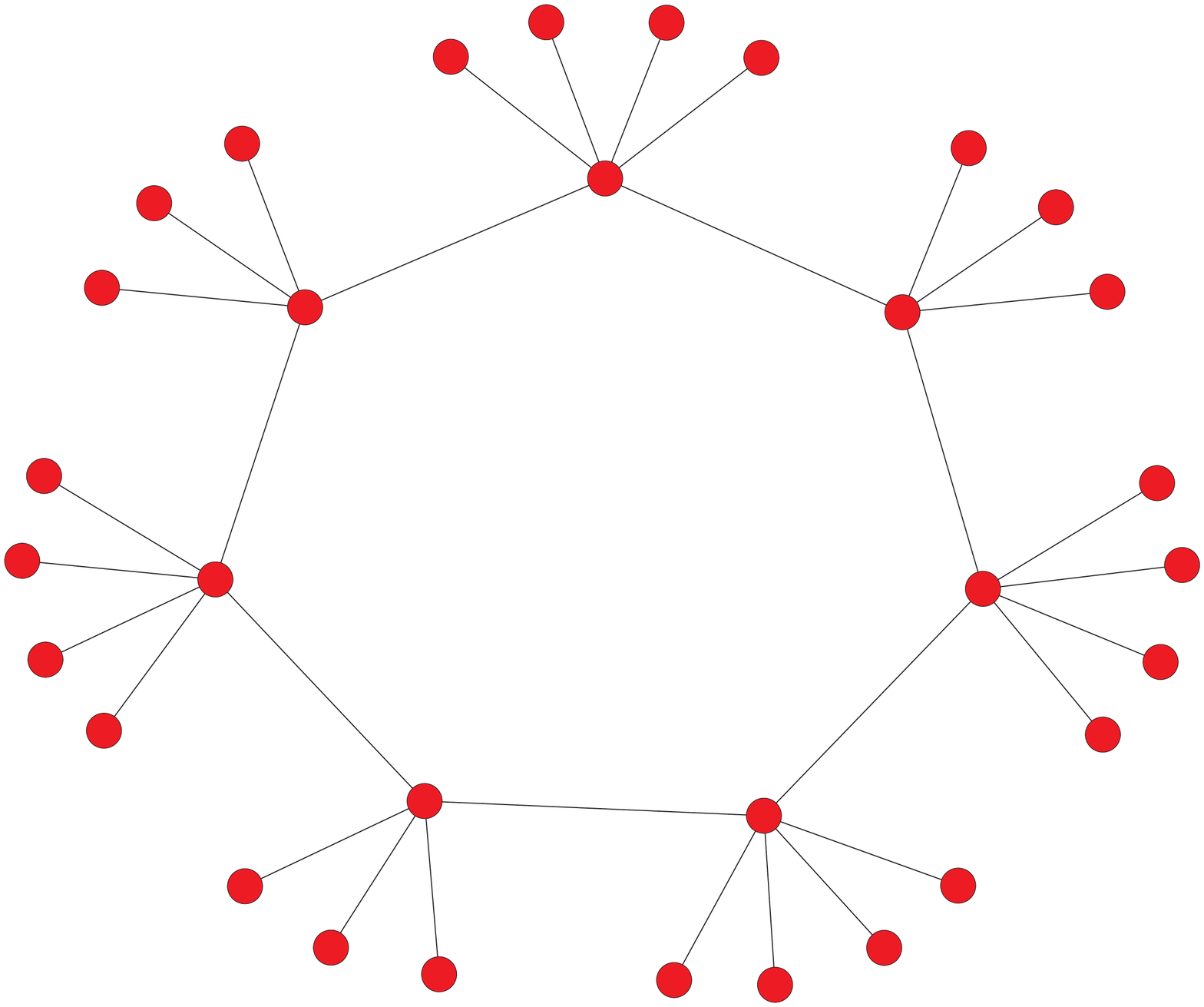}\includegraphics*[width=0.25\columnwidth]{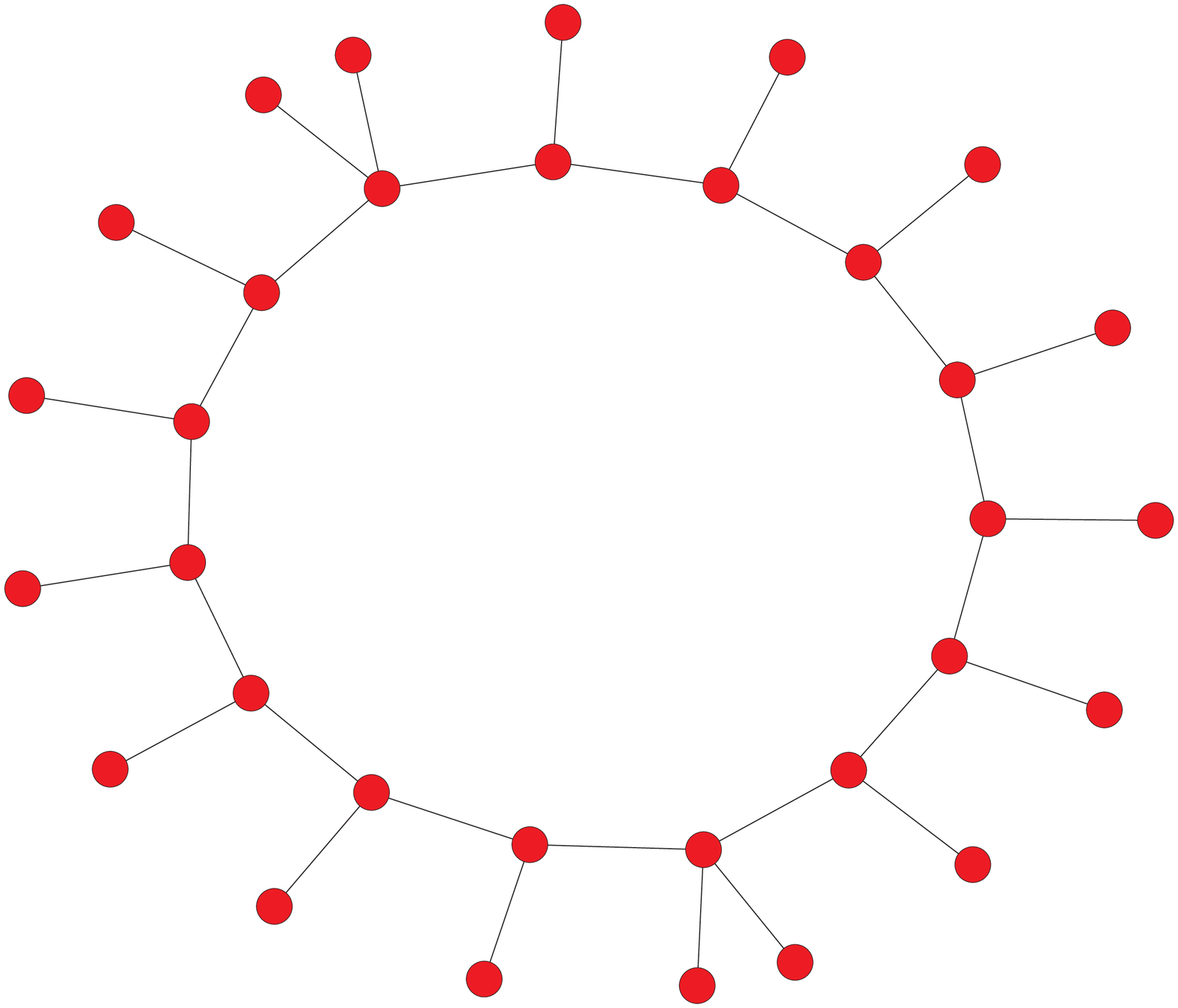}}
\centerline{(a)\hspace{0.2\columnwidth}(b)\hspace{0.2\columnwidth}(c)\hspace{0.2\columnwidth}(d)}
\caption{Optimal topologies for networks with $S=32$ nodes, $L=32$
links and global knowledge. (a) $\rho=0.010$. (b) $\rho=0.020$. (c)
$\rho=0.050$. (d) $\rho=0.080$. In this case of global knowledge, the
transition from centralization to decentralization seems smooth.}
\label{glob}
\end{figure}

Beyond the existence of both centralized and decentralized optimal
networks, it is significant that the transition from one sort
of networks to the other is abrupt, meaning that there are no
intermediate optimal structures between total centralization and total
decentralization. As already mentioned, this property is shared by the
model networks in the previous section. Our explanation of this fact
is the following. Since we are considering (in both the present and
the last sections) local knowledge of the network topology, centered
star-like configurations are extremely efficient in searching
destinations and thus minimizing the effective distance between
nodes. This explains that stars are optimal for a wide range of values
of $\rho$, until the central node (or nodes) becomes congested. At
this point, structures similar to stars will have the same problem and
will be much worse regarding search; at this point, the only
alternative is something completely decentralized, where the absence
of congestion can compensate the dramatic increase in the effective
distance between nodes. If this explanation is correct, one should be
able to obtain a smooth transition from centralization to
decentralization by considering global knowledge of the network, in
such a way that the average effective distance (that in this case
coincides with the average path length) is not much larger in an
arbitrary network than in the star. Although we do not have extensive
simulations in this case, Fig. \ref{glob} shows that there is some
evidence to think that this is indeed the case.

\section{Summary}
We have presented some results concerning search and congestion in networks.
By defining a communication model we have been able to cope with the problems of search
and congestion simultaneously. For a hierarchical lattice some analytical results are found, by
exploiting the symmetry properties of the network. For complex networks, this is not the case, and
computational optimization to look for the best structures is required.
On the one hand, for model networks where short-range, long-range, random and preferential connections
are mixed we find that network that perform well for very low load become easily congested
when the load is increased. On the other hand, when searching for optimal structures in a general scenario
there is a clear transition from star-like centralized structures to homogeneous decentralized ones.

\section*{Acknowledgments}
This work has been supported by DGES of the Spanish Government, Grants No. PPQ2001-1519, No. BFM2000-0626,
No. BEC2000-1029, and No. BEC2001-0980, and EC-FET Open Project No. IST-2001-33555.


%

\end{document}